\documentclass[useAMS,referee]{biom}
\input{psfig}
\usepackage{graphicx}
\usepackage{lineno}
\usepackage{float}
\usepackage{framed}
\usepackage{bm}
\usepackage{Verbatim}
\def\bSig\mathbf{\Sigma}

\floatstyle{plain}
\floatname{panel}{Panel}
\newfloat{panel}{h}{txt}

\title[SCR with partial identity]{
Spatial Capture-recapture with Partial Identity
}

\author
{J. Andrew Royle\emailx{aroyle@usgs.gov} \\
USGS Patuxent Wildlife Research Center, Laurel, MD USA
}

\begin{document}


\date{{\it Received xyz} 2015. {\it Revised xyz} 2015.  {\it
Accepted xyz} 2015.}



\volume{xy}
\pubyear{200x}
\artmonth{May}


\doi{10.1111/j.1541-0420.2005.00454.x}


\label{firstpage}


\begin{abstract}
  I develop an inference framework for spatial capture-recapture data
  when two methods are used in which individuality cannot generally be
  reconciled between the two methods.  A special case occurs in camera
  trapping when left-side (method 1) and right-side (method 2) photos
  are obtained but not simultaneously.  I specify a
  spatially explicit capture-recapture model for the latent
  ``perfect'' data set which is conditioned on known identity of
  individuals between methods. The identity variable which 
  associates individuals of the two data sets is regarded as an unknown in the
  model and I propose a Bayesian
  analysis strategy for the model in which the identity variable is
  updated using a Metropolis algorithm.  The work extends
  previous efforts to deal with imperfect identity information by recognizing that
  there is information about individuality in the spatial juxtaposition
  of captures with encounter devices (traps). Thus, individual records obtained by both sampling
  methods that are in close proximity are more likely to be the same
  individual than individuals that are not in close proximity. The
  model proposed here formalizes the integration of  spatial
  proximity information into models for the probabilistic  
determination of individuality using
  spatially explicit capture-recapture models.

\end{abstract}
%



\maketitle


%

\section{Introduction}
\label{sec.intro}

\begin{figure}[ht]
\centering
\includegraphics[height=3.5in,width=3.5in]{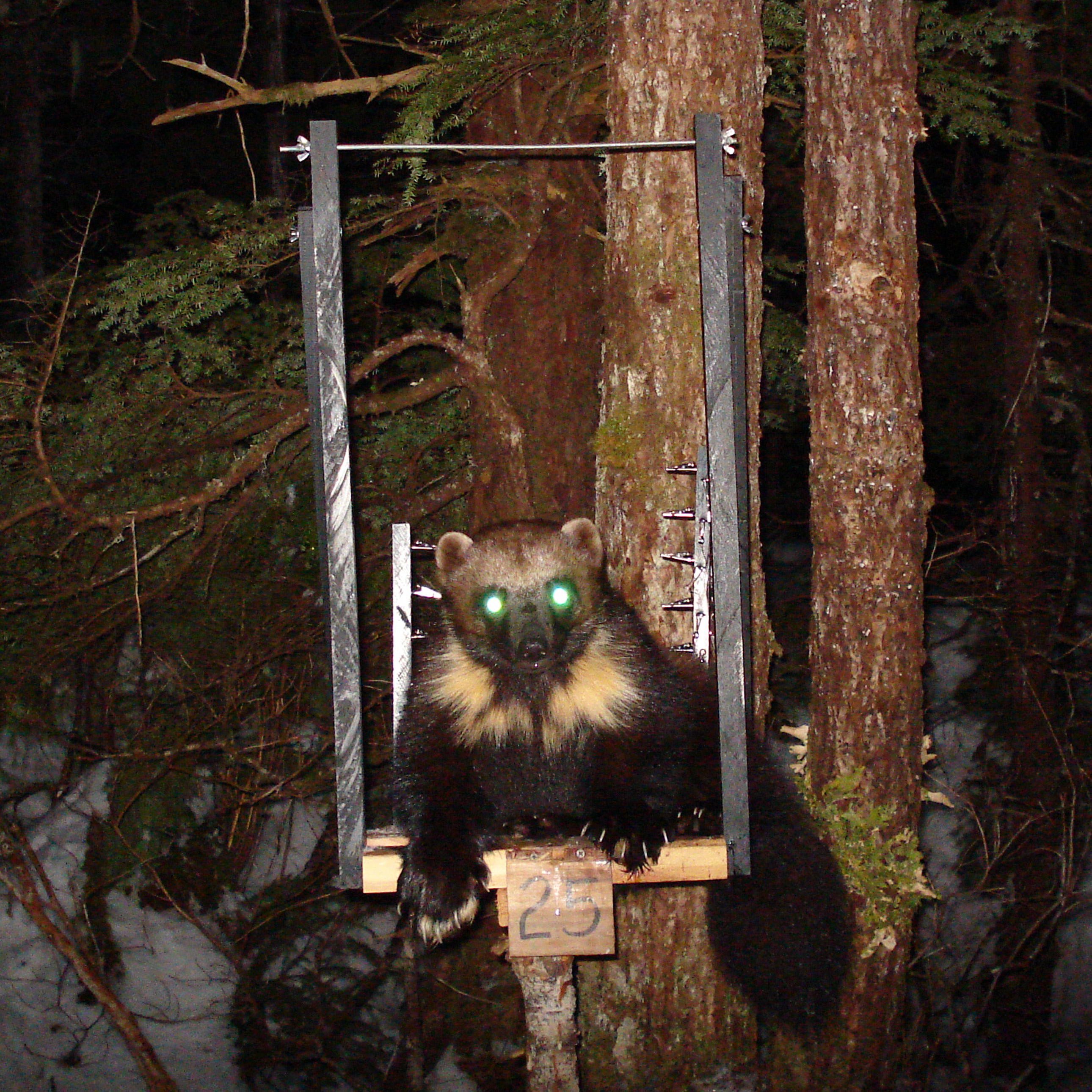}
\caption{
Female  wolverine, Southeast Alaska,  being simultaneously photographed and it's hair
sampled by an alligator clip set-up
(photo credit: A. Magoun, taken from
Magoun et al. 2011a). Wolverines can be uniquely identified by the bib
pattern on their chest (Magoun et al. 2011b).
}
\label{fig.wolverine}
\end{figure}

In this paper, I consider inference for capture-recapture models in which individual
encounter histories are obtained by 2 types of sampling methods which
may not be reconcilable. For example, by camera traps and DNA from
hair snares (Fig. \ref{fig.wolverine}) or camera traps and scat
surveys (Gopalaswamy et al. 2012a).  Each method will, by itself, lead
to individuality of the samples but the methods will not, in general,
lead to the ability to reconcile individuals among themselves.  A
special case occurs when only camera trapping is used, but incomplete
identity is obtained for some individuals, for example right or left
flank only (McClintock et al. 2013; Bonner and Holmberg 2013).  In
camera trapping studies typically photos of both right and left sides
of individuals are needed to produce individual identity (Karanth
1995, O'Connell et al. 2010). However, in practice, sometimes only a
single side is photographed and traditional applications of
capture-recapture based on camera trapping data have discarded these
photos unless subsequent simultaneous photos were obtained.  Camera
trapping by itself can be thought of as a two method sampling problem
with the methods being ``left side camera trap'' and ``right side
camera trap'' while, in some cases, we might have records of both
sides.  A case of special importance is that in which there is a
single camera at every site. In this case, {\it no} reconciliation to
unique individuals is possible from the raw data. However, the
potential to conduct camera trapping studies using only a single
camera is extremely appealing due to the expense of equipment and the
possibility of increasing the number of spatial sampling locations
which is critical to spatial inference problems such as modeling
resource selection (Royle et al. 2013) or landscape connectivity
(Sutherland et al. 2015). Therefore maximizing the statistical
efficiency of information from single trap studies is of substantial 
practical interest.

I propose a model for reconciliation of identity from multiple sample
devices in which true identity is a latent variable.  I provide an
estimation framework using MCMC in which the individual identity of
each encounter history is characterized by Monte Carlo sampling from
the posterior distribution.  The formulation of the model is
facilitated by the observation that, conditional on the true identity,
the model reduces to a basic spatial capture-recapture model for which
MCMC can be implemented directly (Royle et al. 2009, Gopalaswamy et
al. 2012b, Royle et al. 2014, ch. 17), but with a slightly distinct
observation model accounting for multiple observation devices.  
The Metropolis-Hastings algorithm is used to update the latent identity variables.

McClintock et al. (2013) and Bonner and Holmberg (2013) considered
precisely the same conceptual problem I consider here, using the
latent multinomial model of Link et al (2010), but they considered
classical {\it non-spatial} capture-recapture models. A key
consideration in the problem of determining individuality is that
there is direct information about individuality in the {\it spatial
  arrangement of captures}.  For example, if nearby cameras capture
photos of a right and left flank then those are more likely to belong
to the same individual than if those photos were obtained by cameras
far apart.  So far this has not been accounted for in the previous
work on misclassification in capture-recapture (Wright et al. 2009,
Link et al. 2010, McClintock et al 2013, Bonner and Holmberg 2013;
McClintock et al. 2014), although it has been recognized as being
relevant in the combination of data from marked and unmarked
individuals (Chandler and Royle 2013, Chandler and Clark 2014) and
related mark-resight models (Sollmann et al 2013a; Sollmann et al
2013b).  Therefore, in this paper I consider the unknown
individuality in camera trapping or multi-method sampling in the
context of spatially explicit capture-recapture models (Efford 2004;
Borchers and Efford 2008; Royle et al. 2014).  In addition to the
formulation of the model as a spatially explicit model, so that
information arising from spatial location of encounters can be integrated into the
determination of identity, the other element of this paper that
deviates from most existing treatments of the latent multinomial model
is the use of data augmentation (Royle et al. 2007, Royle and Dorazio
2012) to deal with unknown $N$.  Using this formulation of the model
yields a convenient treatment of the unknown population size parameter
$N$ under which the observed data set is augmented to a maximum size
of $M$ with all-zero encounter histories and the model is reformulated
as a zero-inflated version of the known $N$ model. The parameter $N$
is replaced in the model by a set of data augmentation
(zero-inflation) variables $z_i; i=1,\ldots,M$.  This is distinct from
the treatments by Wright et al. (2009) who used reversible jump MCMC
and Link et al. (2010) (and McClintock et al. 2013 and other
applications) whose latent multinomial formulation can accommodate
unknown $N$ by including the unobserved encounter frequency in the
latent multinomial vector. However, that approach does not accommodate
individual level effects, and spatial capture-recapture models as a
class involve a latent individual variable ${\bf s}_i$ the activity or
home range center of individual $i$. The one application of Link et
al. (2010) using data augmentation to account for unknown $N$ is
McClintock et al. (2014) who model individual heterogeneity.  But
their model was not spatially explicit and therefore does not use the
spatial information to inform individuality.

The paper is organized as follows. In the following section I
introduce multi-device observation models that potentially account for
the dependent functioning of two sampling methods. In sec.
\ref{sec.latentdata} I define the 'ideal' data that arise under this
observation model. That is, the data which are produced if individual
identity were known across sampling methods. I also define the
observable data and how the observable data are related to the ideal
data.  In sec. \ref{sec.mcmc} I describe analysis of the model by
posterior simulation using standard Markov chain Monte Carlo methods.
I analyze a simulated data set in sec. \ref{sec.data} to explore
the assignment of identity for a particular case.  In sec.
\ref{sec.simulation} I conduct a simulation study to evaluate the
performance of the Bayesian estimator of population size and encounter
model parameters.  Finally I discuss extensions and the general
relevance of the method in sec. \ref{sec.discuss}.

\section{Spatial encounter models}
\label{sec.multistate}

One of the key ideas developed in this paper is that information about
individuality is available from the spatial pattern of encounters.
For example, two irreconcilable photos are less likely to be the same
individual as the distance between their encounter locations increases
and {\it vice versa}.  Therefore, integrating a model that reconciles
incomplete information on individuality with spatial information about
encounter location requires a modeling framework that accommodates
both types of information.  Spatial capture-recapture models (Borchers
and Efford 2008; Royle et al. 2014) provide this framework.

In SCR models the probability of an individual being captured depends
on both trap location ${\bf x}$ and the spatial location of an
individual's home range, embodied by the activity or home range center
${\bf s}_{i}$ for each individual in the population
$i=1,2,\ldots,N$. The activity centers are regarded as a realization
of a point process and treated in the model as a set of latent
variables (Borchers and Efford 2008, Royle and Young 2008). A typical
model to describe the encounter probability of individuals as a
function of trap location and activity center location is the hazard
model having the form
\[
 p({\bf s}, {\bf x}) = 1-\exp(-h({\bf s}, {\bf x}))
\]
with, for example, $h({\bf s},{\bf x}) = p_0 \exp( -||{\bf s} - {\bf
  x}||^2 /(2\sigma^2) )$ where $p_0$ and $\sigma$ are parameters to be
estimated. Dozens of other models are possible, although cataloging
them is not relevant to anything here.

The activity centers ${\bf s}_{i}$ are not observable, but they are
informed by the pattern of encounters (or non-encounters) of each
individual. The customary assumption is that ${\bf s}_{i} \sim
\mbox{Uniform}({\cal S})$ where ${\cal S}$, the state-space of the
random variables ${\bf s}$, is a region specified in the vicinity of
the traps. Typically the extent and configuration are not important as
long as it is sufficiently large so that encounter probability is near
0 for individuals near the edge (e.g., see Royle et al. 2014,
sec. 6.4.1).  In the 'full likelihood' formulation of the model which
is specified in terms of the population size parameter $N$, being the
number of individuals inhabiting ${\cal S}$, population density is a
derived parameter: $D = N/||{\cal S}||$.

Standard SCR models assume binary or frequency encounter events
governed by the encounter probability model parameterized in terms of
distance between trap and individual activity center. For multi-device
models we require extending the ordinary encounter probability model
to describe more complex encounter events.  Two
possibilities are discussed subsequently.

\subsection{Independent hazards model}

For two devices collocated at point ${\bf x}$ and that function
independently it would make sense to assume an independent hazards
model in which the probability of detection in device $m$ is:
\begin{equation}
 p^{m}({\bf s}, {\bf x}) = 1-\exp(-h^{m}({\bf s},{\bf x}))
\label{eq.hazard}
\end{equation}
I assume
\[
h^{m}({\bf s},{\bf x}) = \lambda_{0} \exp( -\frac{
||{\bf s} - {\bf  x}||^2}{2\sigma^2})
\]
The independent hazards model is sensible in the case of single camera
traps capturing either the left or right flank of individuals.
Further, in this case, it is reasonable to assume $h^{l}({\bf s},{\bf
  x})) =h^{r}({\bf s},{\bf x})) \equiv h({\bf s},{\bf x})$ because it
is the same physical device capturing one side or the other depending
on the orientation of the animal relative to the camera for each visit
to the camera trap location.  Due to our interest in the single camera
trap design I focus on the independent hazards model in the rest of
this paper although briefly discuss an alternative model
 in the next section.
The independent hazards model may also be sensible if two
distinct methods are used such as if camera trapping is used in
conjunction with localized searching for animal scat (e.g., see
Gopalaswamy et al. 2012a) although in this case the devices will not
usually be co-located which is unimportant. 

Under the independent hazards model, the perfect data, that is if you
know identity of individuals, are the {\it paired} binomial frequencies 
${\bf  y}_{ij} = (y_{ij}^{(l)}, y_{ij}^{(r)})$ for each individual
$i$ and trap $j$. This is simply a bivariate version of the standard
encounter frequency data which are analyzed by SCR models.

\subsection{Multi-state observation model}

In general, the independence assumption may not be reasonable. 
For such cases a more general model is required that will account for
dependence of the devices.

Let $p_{i}({\bf s}_i, {\bf x})$ be the probability of capture for
individual $i$ at some trap ${\bf x}$ and then, conditional on
capture, there are 3 possible encounter states (for sampling by 2
devices): the animal can be captured by device A only, device B only
or simultaneously by both devices.  Define the conditional encounter
probabilities for each possible encounter state by the vector
\[
{\bm \pi}_{c} = (\pi_{A},  \pi_{B},  (1-\pi_{A} - \pi_{B}))
\]
and then the unconditional probabilities are 
\[
{\bm \pi} = ( p_{i} \pi_{A},  p_{i} \pi_{B},  p_{i} (1-\pi_{A} -
\pi_{B}), 1-p_{i} )
\]
If the devices are the same type (e.g., both cameras) then we
constrain $\pi_{A} = \pi_{B}$. This is the case for single camera trap
situations and, in that case, the state ``captured by both devices''
is not possible. In this case the multi-state observation model
reduces to the model of the previous section. A sensible
parameterization of the cell probabilities for camera trapping with 2
cameras simultaneously would be to define $\pi_{AB}$ as the
probability of obtaining photos of both sides simultaneously and then
$(1-\pi_{AB})/2$ as the probability of obtaining only one side or
the other.

Non-independence, such as modeled by the conditional-on-capture
multi-state model,  will generally result when devices are collocated
because encounter in one device should lead to a higher probability of
encounter in the 2nd device because the individual is/has already
visited the site on that sample occasion.

\section{Formulation of the model in terms of latent data}
\label{sec.latentdata}

For sampling on $K$ sample occasions (usually nights in a camera
trapping study) the observed bivariate observation for an individual
$i$ of known identity in trap $j$ is ${\bf y}_{ij} = (y_{ij}^{(l)},
y_{ij}^{(r)})$, the frequencies of each type of encounter. For single
camera studies this is the $2 \times 1$ vector of left and right
capture frequencies out of $K$ occasions, and these capture
frequencies are binomial outcomes with parameter $p^{m}({\bf s}, {\bf
  x})$ according to Eq. \ref{eq.hazard}.  A key point of the
subsequent development of things is that the paired data ${\bf y}_{ij}
= (y_{ij}^{(l)}, y_{ij}^{(r)})$ are latent when left and right sides
cannot be perfectly reconciled.  Instead we observe two data sets of
left and right flank encounters having arbitrary (unreconciled, or 'scrambled') row
order. These two data sets are linked to the perfect data ${\bf
  y}_{ij}$ by a reordering of the rows of one or the other of the data
sets. As a simple
illustration Table \ref{tab.one} shows a hypothetical data set that could be observed
based on 4 traps with coordinates (1,2), (1,1), (2,2), (2,1)
respectively and only $K=1$ sample occasion.
\begin{table}[h!]
\centering
\caption{A typical encounter history data set constructed from  
left and right-sided photos with no ability to reconcile the two.
For each data set the rows index individual but  {\it
  not}  across data sets. }
\begin{tabular}{c|cccc|cccc} \hline
     & \multicolumn{4}{c|}{Left data set} & \multicolumn{4}{c}{Right
       data set} \\
trap & 1 & 2 &  3 &  4 &     1 & 2 &  3 &  4 \\ \hline
     & 0 & 1 &  0 &  0 &     1 & 0 &  0 &  0 \\
     & 0 & 0 &  1 &  0 &     0 & 0 &  0 &  1\\
     & 1 & 1 &  0 &  0 &     1 & 0 &  0 &  0\\
     & 0 & 0 &  0 &  1 &       &   &    &    \\
\end{tabular}
\label{tab.one}
\end{table}
These data could represent 7 different individuals or there could be 4
individuals with each row on the right reconciled to one on the left,
we can't know without further information.  However, if we knew how to
match the right encounter histories with the left encounter histories
then we can create our perfect data set ${\bf y}_{ij}$ being the
matched left- and right- encounter frequencies for each individual.
In the sample data given in Table \ref{tab.one}, if we knew that the
rows were in individual order then the perfect data for individual
$i=1$ (row 1), i.e., the left and right encounters ${\bf y}_{1j} =
(y_{1j}^{(l)}, y_{1j}^{(r)})$, are $(0,1)$, $(1,0)$, $(0,0)$, and
$(0,0)$ for the 4 traps. The perfect data can only be constructed
conditional on the ID of the right-side data.  Of course, in practice,
we don't know which rows in the right match with rows in the
left. Lacking specific information on individuals to make this
linkage, one might wonder how it is possible to associate right
encounter histories with left encounter histories. The answer is that
such information comes from the configuration of traps and encounters
of individuals in traps. In our small example, not knowing the true
identity of individuals, we observe a left side individual captured in
trap 4 and also a right side individual captured in trap 4. It
stands to reason that the right side individual is more likely to be
the same individual as left side individual 2 than left side
individual 1 which was only caught in trap 2. Of course to start
making specific probability statements about this we need to be
spatially explicit about our model for encounter probability and the 
distribution of individuals in space. I do this in the next
section.

To distinguish the left and right data sets we'll use the notation
${\bf Yl}$ and ${\bf Yr}$ to be $M \times J$ matrices of encounter
frequencies. Under data augmentation these are of the same dimension
but the rows of the two observed matrices are not reconciled to
individual and therefore do not correspond to the same individual
except by random chance. Despite this, I'll retain the use of a single
subscript $i$ to index rows. Thus, ${\bf Yl}_{i} = (y_{i1}^{(l)},
y_{i2}^{(l)}, \ldots, y_{iJ}^{(l)})$ and ${\bf Yr}_{i} =
(y_{i1}^{(r)}, y_{i2}^{(r)}, \ldots, y_{iJ}^{(r)})$ are the left and
right encounter frequencies, respectively, in each trap
$j=1,2,\ldots,J$. For $K=1$ these are vectors of binary encounters. If
$K>1$ then elements are binomial frequencies based on a sample size of
$K$.

\subsection{Modeling the left-side data set}

To formalize the inference framework I first formulate the problem as
an ordinary SCR model for the left-sided encounter histories and
define the true identity of each individual in the population to be
the row-order of the left-side data set. So in the above example
define individuals 1-4 to be the rows of the left side encounter
history matrix in whatever arbitrary order they are assembled by the
data collector. Note: we could define things vice versa, individuality
defined by the right side data set, but it makes sense to define the
identity based on the larger of the two data sets so there are fewer
unknown IDs to match up (see below).  Henceforth I assert that left is
always the larger data set. If this is not the case for a particular
data set, then we just relabel the data sets.

The appropriate model for the left-side data is a variation of the
ordinary SCR model which has been described in a number of places
including Borchers and Efford (2008) and Royle et al. (2014,
ch. 5). I briefly describe the model here noting that the
full likelihood formulation from Royle et al. (2014) is adopted and not the
conditional likelihood formulation of Borchers and Efford (2008)
although one could formulate the model either way. The left data are
the $n_{left} \times J$ matrix of individual and trap specific
encounter frequencies (out of $K$ nights of sampling). These are
binomial random variables with index $K$ and probability $p({\bf
  s}_{i}, {\bf x}_{j})$ where ${\bf x}_{j}$ is the location of trap
$j$ and ${\bf s}_{i}$ is the activity center of individual $i$, a
latent variable.

I adopt a Bayesian analysis of this model because that proves
convenient in dealing with the unknown identity of the right-side
encounter histories.  In particular, I use the data augmentation
approach to deal with unknown $N$ (Royle et al. 2007,  Royle and Young 2008, Royle 2009,
Royle and Dorazio 2012)
in which the observed encounter history
matrix is augmented with a large number of all-zero encounter histories bringing
the total size of the data set up to $M$, with $n_{left}$ observed
encounter histories (as in the left column of the Table above) and then
$M-n_{left}$ all-zero encounter histories.  We also introduce a set of
$M$ latent variables $z_i$ which accounts for the zero-inflation of
the augmented data set. These data augmentation variables have a
Bernoulli distribution with parameter $\psi$. Under data augmentation
inference is focused on the parameter $\psi$ instead of population
size $N$, but the two are related in the sense that $N \sim
\mbox{Binomial}(M, \psi)$.  Finally, to do a Bayesian analysis of the
SCR model for the left-side data set we need to introduce the latent
activity centers ${\bf s}_{i}$ which are missing data for all $M$
individuals in the augmented data set. All of these details about
Bayesian analysis of SCR models using data augmentation are found in
Royle et al. (2014) and previously cited papers. 

Under the data augmentation formulation of the model, the model for
${\bf y}_{ij}$ the latent or ideal data (when identity is known), 
is specified conditional on the data augmentation variables $z$:
\[
  y_{ij}^{(l)}|(z_{i}= 1)  \sim \mbox{Binomial}(K;   p({\bf s}_{i}, {\bf x}_{j}) )
\]
and
\[
 \Pr( y_{ij}^{(l)} = 0 ) = 1  \mbox{ if $z_{i} = 0$}
\]
The last expression just states that if $z_{i} = 0$ then the only
possible encounter event is ``not captured'' which will occur $K$
times at each trap $j=1,2,\ldots,J$.  

Generalizations of the encounter model are achieved directly by
working with the binary encounter events $y_{ijk}$ in which case the
basic observation model is Bernoulli instead of binomial (Borchers and
Efford 2008; Royle et al. 2014). For example, to model a trap-specific
behavioral response we would need to formulate the model in terms of
the Bernoulli components instead of the binomial frequencies.

\subsection{Linking the right-side encounter histories with the
  left-side encounter histories}

Now consider the right-side encounter histories. These 
individuals are of the same population as the left-side individuals so
we will consider that these right-sided individuals must be associated
uniquely with the left-sided individuals in the population. As our
data sets are augmented to include all-0 encounter histories this
means that some right-sided individuals {\it could} be matched with
left-sided individuals that were not captured.  Therefore, each
right-sided individual which also include $M - n_{right}$ all 0
encounter histories must be uniquely matched to one of the left-sided
individuals {\it and} the pair of data augmentation variable and
activity center ($z, {\bf s}$) that goes with the left-sided
individual. To be clear about this: the $z$ and ${\bf s}$ latent
variables are associated with the left-sided population. Their 'order'
or identity will never change -- $z_{i}$ and ${\bf s}_{i}$ belong with
${\bf Yl}_{i}$, always.  However, ${\bf Yr}_{i}$ will {\it not} be
associated with $z_{i}$, ${\bf s}_{i}$ and ${\bf Yl}_{i}$ in the
arbitrary order by which data sets are assembled and so the
``labeling'' of rows of ${\bf Yr}$ is an unknown parameter of the
model.

Conceptually we make this linkage by introducing a latent ID variable,
${\bf ID} = (ID_{1}, ID_{2}, \ldots, ID_{M})$ where $ID_{i}$ is the
true identity of right-sided individual $i$. That is, $ID_{i}$
indicates which left-sided individual right-sided individual $i$
belongs to.  Formulated in this way we can develop an MCMC algorithm
for this problem that treats the ID variables as latent
variables. Conditional on the $ID$ variables then, we reorder the rows
of ${\bf Yr}$, then recombine ${\bf Yl}$ and ${\bf Yr}$ into the
perfect data set which {\it is} in individual order. Then, given the
perfect data set, we can sample each parameter of the model using
standard MCMC methods (Royle et al. 2014, ch. 17). The prior
distribution for the ID variables under simple random sampling without
replacement of the left-side IDs, from the population of $M$ augmented
individuals, is $\Pr(ID_{i}) = n_{right}/M$, the standard result for
sampling without replacement from a finite population, which doesn't
depend on the parameters of the model (note that data augmentation
replaces the unknown population size $N$ with the fixed dimension $M$,
see Royle et al. 2007).

Conditional on the vector of ID variables, ${\bf ID}$, denote the
re-ordered right-side data set by ${\bf Yr}^{*}$ having elements
$y_{ij}^{(r*)}$ where now the $i$ subscript is ordered consistent with
left-side observations $y_{ij}^{l}$.  Then the observation model is
the same as the left-sided encounter frequencies:
\[
  y_{ij}^{(r*)}|(z_{i}= 1)  \sim \mbox{Binomial}(K;   p({\bf s}_{i}, {\bf x}_{j}) )
\]
and
\[
 \Pr( y_{ij}^{(r*)} = 0 ) = 1  \mbox{ if $z_{i} = 0$}
\]

\section{Estimation by MCMC sampling from the posterior distribution}
\label{sec.mcmc}

Essentially the model described above can be regarded  as a model for
perfect SCR data where the perfect SCR data are linked
deterministically to the observable data, through a latent ID variable
which simply reorders one part (the right-sided encounter histories)
of the data set.  As such, to analyze the model by MCMC we need to do
MCMC analysis of the model for the perfect SCR data and then add a
Metropolis step to update the ID variables. An overview of the algorithm is described
as follows, and a specific implementation in the R language is given
in the appendix. I don't give much detail here because there is
nothing special about the algorithm -- it is plain vanilla
Metropolis-Hastings combined with Gibbs sampling, applied to the basic
SCR model.

\begin{itemize}
\item[0.] Augment the ${\bf Yl}$ and ${\bf Yr}$ data sets with
  all-zero encounter histories to bring them both to dimension $M
  \times J$. 
\item[1.] Initialize all parameters, $p_0$, $\sigma$, $\psi$ for 
  the ``single camera trap'' model, and the $M$ data augmentation
  variables $z_i$ and the $M$ ID variables. 
\item[2.] Given the current value of the ID variables 
 we simply re-order the rows of ${\bf Yr}$ so they are in
 individual-order (consistent with the ${\bf Yl}$ data set). We call
 this re-ordered data set ${\bf Yr}^{*}$. 
\item[3.] Given the ${\bf Yl}$ and ${\bf Yr}^{*}$ data sets,
encounter histories we can construct the reconciled individual and trap specific
frequencies ${\bf y}_{ij}=(y_{ij}^{(l)}, y_{ij}^{(r*)})$.
\item[4.] Conditional on the current perfect data set, we can update
  SCR model parameters $\sigma$ and  $p_0$
conditional on the current encounter
history configuration using standard Metropolis-Hastings.
\item[5.] We update each ID variable that is unknown using a standard
  Metropolis step. We note that each left-side individual can only
  have a single right-side individual associated with it. Thus we have
  to sample IDs without replacement. However, this is conveniently
  done sequentially, one right-side individual at a time, by swapping
  ID's between a pair of right-side individuals. That is, for each
  right-side individual, we pick a different right-side individual,
  and we {\it swap} their $ID_{i}$ values to produce a candidate ID
  vector ${\bf ID}^{*}$.  Given the candidate ID vector, which is to
  say the ID vector having made a potential swap, we have to recompute
  the perfect data set and then the acceptance ratio for the candidate
  value of $ID_{i}$ is the likelihood ratio of the perfect data given
  the candidate value of $ID_{i}$ to the current value of $ID_{i}$. We
  used a distance neighborhood to look for potential IDs to swap and
  the result is a non-symmetric proposal distribution which must be
  accounted for by the usual Metropolis acceptance probability. We can
  cycle through the right-side encounter histories for a set number of
  swaps (e.g., 10, 50 or 100). Too few doesn't update enough IDs
  during any MCMC iteration but too many is inefficient because it
  updates many of the IDs multiple times.
 \item[6.] We update each $z_{i}$ variable as in an 
ordinary SCR model (see Royle et al. 2014, chapt. 17). 
\item[7.] We update each ${\bf s}_{i}$ as done in an ordinary SCR model
  (see Royle et al. 2014, chapt. 17). 
\end{itemize}

An R script for simulating data and for fitting the model to simulated
data is provided in an R package located at
\mbox{\tt https://sites.google.com/site/spatialcapturerecapture/misc/1sided}.
There are no important technical considerations in implementing this
algorithm for the single-flank camera trapping situation. However, it is
extremely helpful to choose initial values for $ID_{i}$ which
associate each right-sided individual with a left-sided individual
that is not to far away, or else the possibility exists that, for the
current (or starting) value of $\sigma$ you could have an encounter
that has 0 probability of happening.  The R code includes a function
that will sort through $ID$ matches to find a set of matches which
minimizes the total distance of captures between matched pairs.

\subsection{Known identity individuals}

In some studies, even if the study uses only a single camera trap per
station, some of the individual ID's may be known. For example, it is
common to conduct a telemetry study that is often done simultaneous to
a capture-recapture study. The individuals physically captured for
telemetry will be known perfectly and thus when they are detected by
the capture recapture study their identity will be known.  In the MCMC
algorithm outlined above, by convention, we organize the left and
right data sets so the first $n_{known}$ rows correspond to these
``known ID'' individuals.
In the case of an independent telemetry study which
produces the first sample of captured individuals it may be reasonable
to assume that the known-ID individuals are selected randomly from the
population of $N$ individuals on the state-space ${\cal S}$ (Chandler
and Royle 2013; Sollmann
et al. 2013a) and that the number of known-ID individuals is known
perfectly. Under these two assumptions, the encounter observations for
the known-ID individuals are regarded as binomial encounter
frequencies on par with the individuals observed in the
capture-recapture study but they are not used to provide information
about $N$. Then, the SCR model provides an estimate of the number of
individuals whose true ID is unknown.  Note also that some known-ID
individuals may not show up in the capture-recapture study and, in
this case, their all-zero encounter history is {\it observed}.

\subsection{The heuristic estimator}
\label{sec.heuristic}

An intuitive procedure for dealing with left- and right-side encounter
histories is to treat them as independent samples from the sample
population having a single $N$, $p_0$ and $\sigma$ parameter. This is
easy to do using the data augmentation approach by introducing 2 sets
of latent data augmentation variables say $z_{left,i}$ and
$z_{right,i}$ which are independent of one another but have a common
$\mbox{Bernoulli}(\psi)$ distribution.  The model in this case is just
that of two independent SCR models and the MCMC algorithm described by
Royle et al. (2014, ch. 17) applies directly (but duplicating the
likelihood part).

We expect that this procedure should yield unbiased estimates of population
size and hence density.  On the other hand there may be a
misspecification of variance because it regards the observed data as
two independent data sets of size $n_{left}$ and $n_{right}$, i.e., on
average twice the number of individuals you actually observe. In
reality there is a non-independence between the encounter histories
that the heuristic estimator is not accounting for. In fact, the model
allowing for ID to be unknown, described in this paper, accommodates
that dependence structure formally and so it {\it is} the model for
non-independence of the left- and right-side encounter histories.
 
\section{Analysis of a data set}
\label{sec.data}

\begin{figure}[h]
\centering
\includegraphics[height=4in,width=4in]{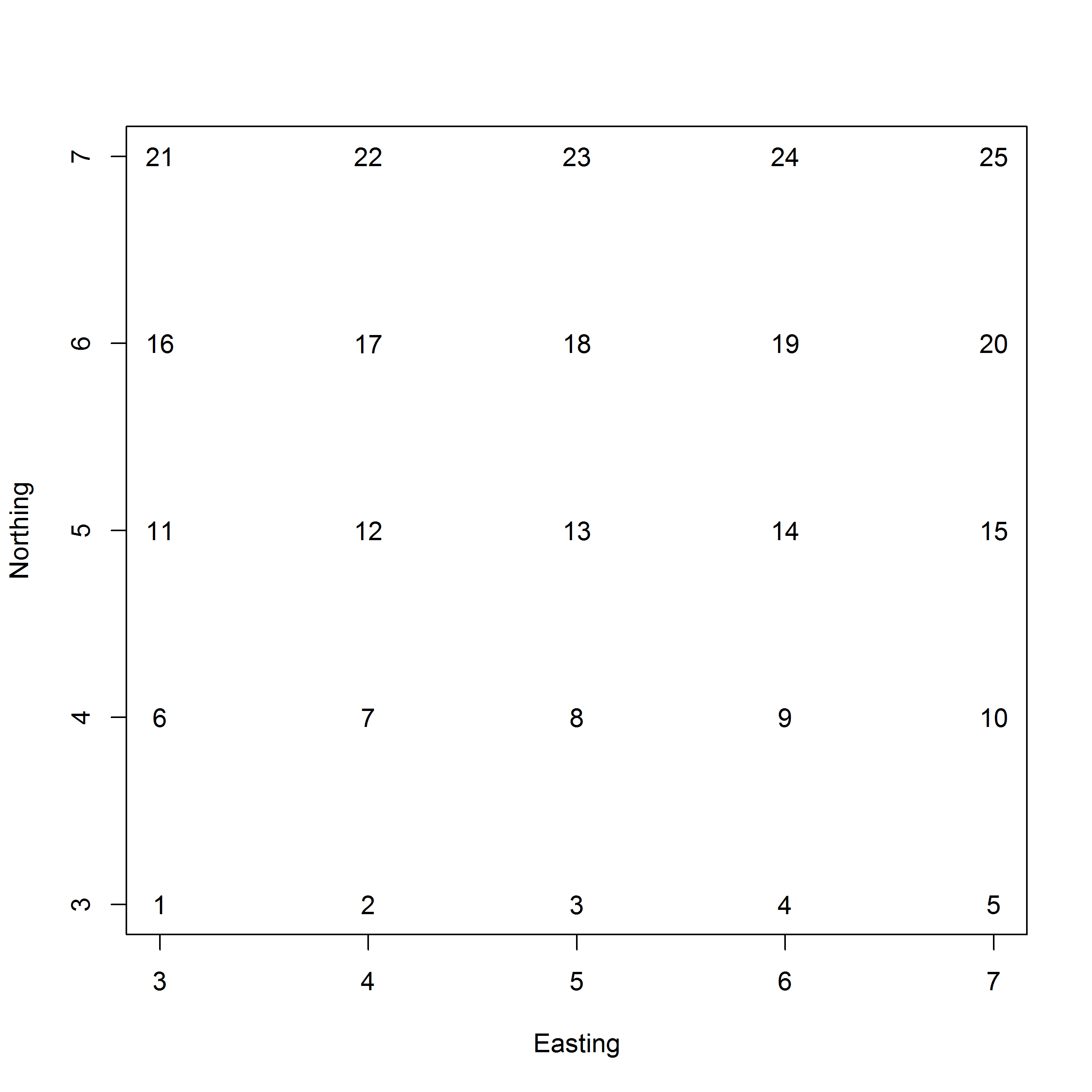}
\caption{
Grid of 25 traps of unit spacing 
used in simulating left-sided and right-sided encounter
history data from a study using a single
camera-trap per site.  Trap identifier is given for reference in
text. 
}
\label{fig.fig1}
\end{figure}

I simulated a data set with $N=60$ individuals exposed to trapping by
a $5\times 5$ grid of single camera traps having unit spacing. 
The arrangement of traps along with trap numbers of identification
purposes is shown in Fig. \ref{fig.fig1}. 
Data were generated using the hazard model having form 
$p({\bf s}, {\bf  x}) = 1-\exp(-h({\bf s},{\bf x}))$ with 
\[
h({\bf s}, {\bf  x}) =
 p_0 \exp( \frac{ - ||{\bf s} - {\bf x}||^{2} }{2 \sigma^2} )
\]
and with $\sigma = 0.5$ and $p_0 = 0.2$.  This generated a data set 
investigated here that has $n_{left} = 28$ left side individuals and
$n_{right} = 24$ individuals. Because the data are simulated, it is known
 that the total number of observed individuals was 30.
Posterior summaries based on 1000 burn-in and 2000 post-burn in
samples are given in Table \ref{tab.posterior}.
\begin{table}[h!]
\centering
\caption{Posterior summaries of model parameters from a simulated
  population of $N=60$ individuals, $p_0 = 0.20$ and $\sigma =
  0.50$. The population was augmented to $M=120$ individuals and
  therefore $\psi = 0.50$. }
\begin{tabular}{c|rrrrrrr} \hline
parameter & mean & SD& 2.5\%& 25\% & 50\% &75\% & 97.5\% \\ \hline \hline
$p_0$     & 0.182& 0.033&  0.126 & 0.160 & 0.177 & 0.204 & 0.254\\
$\sigma$  & 0.517& 0.037&  0.450 & 0.487 & 0.517 & 0.543 & 0.590\\
$\psi$    & 0.452& 0.080&  0.313 & 0.394 & 0.448 & 0.505 & 0.619\\
$N$       &54.006& 8.101& 41.000 &48.000 &53.000 &59.000 &72.000
\end{tabular}
\label{tab.posterior}
\end{table}
We see the posterior of $N$ is not too far from the truth (mean 54), nor are the
posterior means of $\sigma$ and $p_0$ far from the data generating
values for this 
specific realized data set. 

Let us consider using our posterior simulated results for the $ID$
variable to estimate the identity of individual right-side 11 which
was captured 2 times in trap 25 but no other traps.  There were a
number of individuals in the left data set caught in close proximity
to right-side 11 and we're especially interested in whether the model
associates right-side individual 11 with these nearby left-side
individuals.  To investigate this, capture frequencies of the 18
individuals ordered by closeness of average capture location to
individual right-side 11 are shown in Table \ref{tab.guy11}. Only the
frequency of captures in traps 13-25 are shown as the other traps are
distant from trap 25. We see a number of candidates that right-side
11 may belong to. In particular, the left-side individual 16 here was
captured 1 time in each of traps 20 and 25, left-side 21 was captured
once each in trap 24, and so on.  To see how the model evaluates this
information the posterior frequency of ID is shown in the column
``post. Freq.''  We see that in fact the posterior ID of individual
right-side 1 is highly associated with those left-side individuals
captured in its vicinity. Note also that there were 391 posterior
samples where right-side 11 is judged to be a ``new'' individual. That
is, {\it not} one of the 28 captured left-side individuals. In fact,
because we simulated the data and know the true identity in this case
we know that right-side 11 is actually the same as left-side
individual 11 which receives posterior mass 216/2000 and was captured
two times in trap 20 (a neighboring trap). The model seems to be doing
a sensible thing in this instance.

\begin{table}
\centering
\caption{Posterior assignment of right-side individual 11, captured
  twice in trap 25, to various
  left-side individuals. The posterior frequency of an ID match is
  shown in the far right column. For example, 692 posterior samples
  show that right-side individual 11 goes with left-side individual
  16. The true ID in this case has right-side 11 also belonging to
  left-side 11 (note: right and left IDs are arbitrary so this is a
  coincidence). The middle columns of the table are trap
  identifications and the numbers are frequency of capture in each
  trap. 
}
\begin{tabular}{rcccccccccccccr} \hline \hline
leftID & \multicolumn{13}{c}{Trap ID} & post. Freq. \\ \hline
   & 13&14& 15 &16& 17& 18& 19& 20& 21& 22& 23& 24& 25 \\    \hline
16  & 0 & 0 & 0 & 0 & 0 & 0 & 0 & 1 & 0 & 0 & 0 & 0&  1 &692\\
7   & 0 & 0 & 0 & 0 & 0 & 0 & 1 & 0 & 0 & 0 & 0 & 1&  1  &72\\
{\bf 11}&  0 & 0&  0&  0&  0&  0&  0&  2&  0&  0&  0&  0 & 0& 216 \\
20  &0 & 0 & 0 & 0 & 0 & 0 & 0 & 0 & 0 & 0 & 0 & 2 & 0  & 0\\
21  &0 & 0 & 0 & 0 & 0 & 0 & 0 & 0 & 0 & 0 & 0 & 1 & 0 &288\\
22  &0 & 0 & 0 & 0 & 0 & 0 & 0 & 1 & 0 & 0 & 0 & 0 & 0 &342\\
24  &0 & 0 & 0 & 0 & 0 & 0 & 0 & 0 & 0 & 0 & 1 & 0 & 0 &  0\\
27  &0 & 0 & 1 & 0 & 0 & 0 & 0 & 0 & 0 & 0 & 0 & 0 & 0 &  0\\
1   &0 & 0 & 0 & 0 & 0 & 4 & 0 & 0 & 0 & 0 & 3 & 0 & 0 &  0\\
15  &1 & 0 & 0 & 0 & 0 & 0 & 1 & 0 & 0 & 0 & 0 & 0 & 0 &  0\\
12  &0 & 2 & 0 & 0 & 0 & 0 & 0 & 0 & 0 & 0 & 0 & 0 & 0 &  0\\
9   &0 & 0 & 0 & 0 & 0 & 0 & 0 & 0 & 0 & 1 & 2 & 0 & 0 &  0\\
3   &0 & 0 & 0 & 0 & 0 & 0 & 0 & 0 & 0 & 2 & 1 & 0 & 0 &  0\\
4   &0 & 1 & 0 & 0 & 0 & 0 & 0 & 0 & 0 & 0 & 0 & 0 & 0 &  0\\
6   &0 & 0 & 1 & 0 & 0 & 0 & 0 & 0 & 0 & 0 & 0 & 0 & 0 &  0\\
17  &2 & 0 & 0 & 0 & 0 & 0 & 0 & 0 & 0 & 0 & 0 & 0 & 0 &  0\\
19  &0 & 0 & 0 & 0 & 0 & 1 & 0 & 0 & 0 & 0 & 0 & 0 & 0 &  0\\
8   &0 & 0 & 0 & 0 & 2 & 0 & 0 & 0 & 0 & 0 & 0 & 0 & 0 &  0 \\
391  &0 & 0 & 0 & 0 & 0 & 0 & 0 & 0 & 0 & 0 & 0 & 0 & 0 &  391 \\ \hline
\end{tabular}
\label{tab.guy11}
\end{table}

As a 2nd example, consider right-side individual 21 which was caught 1
time in trap 4. The posterior frequency distribution is shown in 
Table \ref{tab.guy21}
where 
the frequencies of encounter in traps 1-12 are shown,  because those are
closest to the capture of individual 21 in trap 4.
Note also that 
 the posterior frequency that right-side individual 21 was {\it not}
associated with any observed left-side individual was 201/2000. In
fact, this individual actually belongs with left-side individual 13
and so the model predicts the correct individuality with probability $0.73$.

\begin{table}
\centering
\caption{Posterior assignment of right-side individual 21, caught 1
  time in trap 4,  to eighteen other individuals captured in close proximity.
 The posterior frequency of an ID match is
  shown in the far right column. For example, 1469 posterior samples
  show that right-side individual 21 goes with left-side individual
  13 which is the true ID of right-side individual 21. 
 The middle columns of the table are trap
  identifications and the numbers are frequency of capture in each
  trap. 
}
\begin{tabular}{rccccccccccccr} \hline \hline
leftID & \multicolumn{12}{c}{Trap ID} & post. Freq. \\ \hline
       & 1& 2 &3 &4 &5 &6 &7 &8 &9 &10 &11 &12 &    \\      \hline
{\bf 13}& 0& 0 &0 &2 &0 &0 &0 &0 &0  &0 & 0 & 0 &1469  \\
26& 0& 0 &0 &0 &1 &0 &0 &0 &0  &0 & 0 & 0 & 326\\
6 & 0& 0 &0 &0 &0 &0 &0 &0 &2  &0 & 0 & 0 &   0\\
4 & 0& 0 &0 &0 &0 &0 &0 &0 &0  &2 & 0 & 0 &   5\\
2 & 0& 0 &1 &0 &0 &0 &0 &2 &0  &0 & 0 & 1 &   0\\
5 & 0& 2 &1 &0 &0 &0 &0 &0 &0  &0 & 0 & 0 &   0\\
12& 0& 0 &0 &0 &0 &0 &0 &0 &0  &0 & 0 & 0 &   0\\
17& 0& 0 &0 &0 &0 &0 &0 &0 &0  &0 & 0 & 0 &   0\\
27& 0& 0 &0 &0 &0 &0 &0 &0 &0  &0 & 0 & 0 &   0\\
14& 0& 0 &0 &0 &0 &0 &1 &0 &0  &0 & 0 & 1 &   0\\
15& 0& 0 &0 &0 &0 &0 &0 &0 &0  &0 & 0 & 0 &   0\\
25& 0& 0 &0 &0 &0 &0 &0 &0 &0  &0 & 0 & 1 &   0\\
19& 0& 0 &0 &0 &0 &0 &0 &0 &0  &0 & 0 & 1 &   0\\
11& 0& 0 &0 &0 &0 &0 &0 &0 &0  &0 & 0 & 0 &   0\\
22& 0& 0 &0 &0 &0 &0 &0 &0 &0  &0 & 0 & 0 &   0\\
28& 0& 0 &0 &0 &0 &1 &0 &0 &0  &0 & 0 & 0 &   0\\
8 & 0& 0 &0 &0 &0 &0 &0 &0 &0  &0 & 1 & 0 &   0\\
1 & 0& 0 &0 &0 &0 &0 &0 &0 &0  &0 & 0 & 0 &   0\\
none & 0& 0 &0 &0 &0 &0 &0 &0 &0  &0 & 0 & 0 &  201 \\ \hline
\end{tabular}
\label{tab.guy21}
\end{table}

We can match up individuals like this all day and it's a lot of fun to
see the model decide which individuals captured on the right-side
should be matched to left-side individuals. It's almost more fun than
the objective of estimating population size.

\section{Simulation study}
\label{sec.simulation}

To evaluate the performance of the SCR model of single-sided camera
trapping, I did a simulation study focused on estimating population
size $N$ using a $5 \times 5$ grid of traps having unit spacing. The
state-space for the SCR model was constructed by buffering the square
trap array by 2 units.  Parameter values were structured according to
a factorial design with 2 levels of $N$ (80,120) two levels of
$\sigma$ (0.70, 0.50) and 2 levels of $p_0$ (0.20, 0.10), for a total
of 8 distinct scenarios.  For each of the 8 parameter settings I
evaluated 4 estimation scenarios: (1) A perfect data situation in
which the true ID is known for all individuals so that left and right
side encounter histories can be reconciled perfectly; (2) 10
individuals have known ID such as captured for telemetry prior to the
capture-recapture study. Thus, the right and left side of these 10
individuals is reconciled perfectly; (3) none of the right-side
individuals have a known ID and therefore ID must be determined using
the model described in this paper; and (4) the heuristic estimator as
described in section \ref{sec.heuristic}.  Thus, a total of 32
distinct estimation scenarios were carried out (8 parameter scenarios
times 4 estimation scenarios).  The first estimation scenario is an
ordinary capture-recapture model but with left- and right- photos
recorded separately and, as a result, maybe occur with unequal
frequency by chance. This is not a typical data structure although
such data could easily be recorded in practice. Here it serves to
provide a baseline to which the remaining cases, which have relatively
less information, can be compared.
Two-hundred (200) data sets were simulated for each of the 32
parameter/estimation scenarios and analyzed by MCMC by sampling 21000
values from the posterior distribution, discarding the first 1000
values as burn-in.  Each MCMC analysis took between 1 and 2 hours of
CPU time, thus about 200-400 CPU hours for each of the 32 scenarios.
Summaries of the posterior distribution were computed for each of the
200 simulated data sets, including the posterior mean and posterior
mode. A frequentist evaluation of the 'sampling error' was computed as
the standard deviation among the posterior summaries for the 200 data
sets. In addition I computed the frequentist coverage of a 95\%
confidence interval for $N$ based on the 2.5 and 97.5
percentiles. Finally I computed the average posterior standard
deviation and the expected posterior mode by pooling the posterior
distributions from all 200 data sets.

The values of the parameters were
chosen by trial-and-error to produce data sets of variable expected sample sizes
of captured individuals.  See Table \ref{tab.ss} for the average sample sizes
of total number individuals and total capture events. 
 The high value of $N$ should generate high
sample sizes of individuals and, as a result, more uncertainty in
individual assignment as there will be more individuals in the
vicinity of each trap. Thus we expect relatively better performance in
the $N=80$ scenarios, which is to say more loss in precision compared
to ``all known'' and relatively better precision compared to the
heuristic estimator. Similarly if we were to increase the number of
traps, holding $N$ constant, we should expect improved performance
(see Chandler and Royle 2013 for a similar phenomenon).
Simulation results for the $N=120$ cases are summarized in Table
\ref{tab.sim1} (the $N=80$ cases are given in the Appendix).

\begin{table}[ht]
\caption{Expected number of unique individuals captured ($n_{left}$,
  $n_{right}$) and total number of capture events ('caps') for each
  simulation scenario. Three table sections represent no known ID
  individuals ($nID=0$), 10 known ID individuals ($nID = 10$) and all
  known ($nID = all$). Note that in the first two groups where
  individuals have unknown ID the data sets are ordered upon
  collection so that the left-side encounter history data set is
  larger, so you see a systematically larger left-side data set. 
}
\begin{tabular}{cc|cccc|cccc}
  \multicolumn{5}{c}{no known ID individuals, $nID = 0$} &  & & & &  \\
            &         & \multicolumn{4}{c}{ N=120 } &  \multicolumn{4}{c}{N=80} \\
$p_0$ & $\sigma$ &  $E(n_{left})$ & caps & $E(n_{right})$ &caps &
$E(n_{left})$ &  caps & $E(n_{right})$ & caps \\ \hline
0.20 & 0.70 & 79.2 &278.5 &75.6 &273.9 &  53.1 &184.7 &50.4 &181.0  \\
0.10 & 0.70 & 65.4 &143.3 &61.4 &137.3 &  44.3 & 96.3 &41.1 & 91.1 \\
0.20 & 0.50 & 58.8 &142.9 &55.5 &137.4 &  40.0 & 96.1 &37.4 & 90.8 \\
0.10 & 0.50 & 45.8 & 74.3 &41.8 & 68.5 &  31.8 & 50.9 &28.8 & 46.0 \\ \hline
\multicolumn{5}{c}{10 known ID individuals, $nID =  10$} &  & & & & \\
            &         & \multicolumn{4}{c}{ N=120 } &
            \multicolumn{4}{c}{N=80} \\ 
$p_0$ & $\sigma$ &  $E(n_{left})$ & caps & $E(n_{right})$ &caps &
$E(n_{left})$ &  caps & $E(n_{right})$ & caps \\ \hline
0.20 & 0.70 &  82.0 &278.5& 78.4 &273.9 & 55.9 &184.7 &53.2 &181.0 \\
0.10 & 0.70 &  69.2 &143.3& 65.1 &137.3 & 48.1 & 96.3 &44.8 & 91.1\\
0.20 & 0.50 &  63.4 &142.9& 60.1 &137.4 & 44.6 & 96.1 &42.0 & 90.8\\
0.10 & 0.50 &  51.1 & 74.3& 47.2 & 68.5 & 37.0 & 50.9 &34.1 & 46.0\\ \hline
\multicolumn{5}{c}{all known ID individuals, $nID =  all$} &  & & & &  \\
           &         & \multicolumn{4}{c}{ N=120 } &
            \multicolumn{4}{c}{N=80} \\ 
$p_0$ & $\sigma$ &  $E(n_{left})$ & caps & $E(n_{right})$ &caps &
$E(n_{left})$ &  caps & $E(n_{right})$ & caps \\ \hline
 0.20  & 0.70 &  87.3& 278.4 & 87.3 &271.4 & 58.3& 185.2 & 58.3 &180.5  \\
 0.10  & 0.70 &  76.0& 145.7 & 76.0 &137.9 & 50.6&  97.0 & 50.6 & 90.7  \\
 0.20  & 0.50 &  65.3& 142.7 & 65.3 &136.4 & 43.9&  96.7 & 43.9 & 91.0 \\ 
 0.10  & 0.50 &  56.2&  74.8 & 56.2 & 69.0 & 36.6&  49.4 & 36.6 & 44.2 \\
\end{tabular}
\label{tab.ss}
\end{table}

\begin{table}[ht]
\caption{Simulation results for the population size parameter $N$
  under the $N=120$ scenarios. For each of the parameter settings
  (cases 1-4), four different estimators are considered: 'nID=0' is that
  in which no individuals have known ID, 'nID=10' and 'nID=all' are 10
  known and all known, respectively, and 'heur' is the heuristic
  estimator which regards left and right side encounter histories as
  independent samples from the posterior distribution.  For each case
  (organized in rows), summaries are the mean of the
  posterior mean and mode over 200 simulated data sets ('mean' and
  'mode' respectively) and the standard deviation of those 200 values
  ('sd'), the average posterior standard deviation ('postSD'), the
  frequentist coverage of a 95\% posterior interval based on the 2.5
  and 97.5 percentiles ('95cover') and the overall posterior mode
  ('pmode').
}
\begin{tabular}{crrrrrrr} 
\multicolumn{2}{c}{Case 1: $\sigma = 0.7, p_0 = 0.20$ } 
                   &       &         &       &        &             &  \\
Model   &  mean   & sd    & mode    &  sd   & postSD & 95cover &pmode \\ \hline
nID=0   & 120.076 & 8.248 & 118.970  & 8.335 & 7.772 & 0.935 & 120\\
nID=10  & 120.064 & 7.847 & 118.953 & 7.872 & 7.405 & 0.950 & 119\\
nID=all & 120.851 & 7.345 & 119.698 & 7.357 & 7.223 & 0.915 & 120\\
heur    & 120.73  & 8.063 & 120.027 & 8.183 & 6.499 & 0.850 &119\\ \hline
\multicolumn{2}{c}{Case 2: $\sigma = 0.7, p_0 = 0.10$ } 
  &       &         &       &        &              &  \\
Model   & mean    & sd     & mode    & sd    & postSD &95cover&  pmode  \\ \hline
nID=0   & 121.844 & 10.491 & 120.235 & 10.354 & 10.314 & 0.930 &  120\\
nID=10  & 121.778 & 10.086 & 120.33 & 10.203 & 9.771 & 0.925 &  120\\
nID=all & 120.866 & 8.931  & 119.562 & 9.008 & 9.228 & 0.940 &  119\\
heur    & 120.768 & 10.269 & 119.534 & 10.264 & 8.976 & 0.915 & 117.5\\ \hline
\multicolumn{2}{c}{Case 3: $\sigma = 0.5, p_0 = 0.20$}
  &       &         &       &        &         &      \\
Model  &  mean  &   sd  & mode    &  sd    & postSD & 95cover&pmode\\ \hline
nID=0  & 121.250 & 11.545 & 119.780 & 11.481 & 11.114 & 0.930 &120 \\
nID=10 & 121.010 & 10.707 & 119.275 & 10.597 & 10.569 & 0.920 & 120\\
nID=all& 121.980 & 10.094 & 120.435 & 9.838 & 10.493 & 0.960  & 122\\
heur   & 122.276 & 10.823 & 121.188 & 10.662 & 9.135 & 0.890  &120.5\\ \hline
\multicolumn{2}{c}{Case 4: $\sigma = 0.5, p_0 = 0.10$ }
  &       &         &       &               &     &  \\
Model  &   mean &  sd    &  mode  &  sd    & postSD & 95cover & pmode \\ \hline
nID=0  & 123.736 & 15.049 & 120.799 & 15.120 & 15.572 & 0.955 &  119\\
nID=10 & 124.041 & 14.566 & 121.135 & 14.599 & 14.737 & 0.955 &  122\\
nID=all& 121.507 & 13.368 & 119.412 & 13.363 & 12.985 & 0.935 &  118\\
heur  & 121.906 & 14.783 & 119.201 & 14.708 & 14.117 & 0.935 &  117.5
\end{tabular}
\label{tab.sim1}
\end{table}

Some salient points of the results presented in Table \ref{tab.sim1} are
summarized here:  1. Main interest is in the basic estimator when we have no known
 ID individuals. That is, the $nID = 0$ case. For this case we see the
posterior mode is usually centered over the correct value (120, 120, 
120, 119) for the 4 cases. The average posterior mean, if used as a
frequentist point estimator, indicates small bias due to the slight
skew of the posterior distribution. The average coverage of 95\%
posterior intervals is about 94\%.   The posterior SD for this case
is usually < 10\% larger compared to the ``all known'' case. However,
for the $\sigma = 0.5$, $p_0 = .10$ case (relatively sparse in terms
of recaptures) we see that the posterior SD is about 20\% worse than
the known ID case. 
2. As the information content increases from 0 known to 10 known to
all known we naturally
see an increase in precision of the posterior mean and mode as  frequentist
point estimators, and a decrease in average posterior SD.
3. The heuristic estimator performs only slightly better than the ``0
known'' case in terms of frequentist SD of the mean and mode as point
estimators.  The coverage of a 95\% confidence interval is only about
90\% when averaged over the 4 cases. 
The estimator is too liberal which may be an expected result due to the
heuristic estimator over-counting the sample size (treating left and
right side data sets as two independent data sets). This shows up in
the smaller posterior standard deviations for the heuristic estimator.

\section{Discussion}
\label{sec.discuss}

I developed a spatial capture-recapture model for dealing with
individual encounter history data in which individuality is
imperfectly observed. The specific motivating example is camera
trapping using a single camera trap. In such cases, only
 left or right side photos are obtained, and there is no
direct information about which right-side encounter history goes with
which left-side encounter history. Even when 2 camera traps are used
at each site (or a subset of sites) sometimes only one flank of the
animal will be identifiable either due to poor image quality or
failure of the device. In practice these one-sided cases have been
ignored unless a subsequent match is produced from 2 simultaneous
images. Thus the model I proposed here allows for more data to be
used from standard 2-camera trap studies and it also allows for the
flexibility of 1-camera per site studies.

The basic model of uncertain identification is similar to Link et
al. (2010) and indeed the Link et al. (2010) idea was applied by
Bonner and Holmberg (2013) for multiple sampling methods and
McClintock et al. (2013) to the problem of bilateral photo
identification. However the  model I described here is distinct in that it is {\it
  spatially explicit} so that it accommodates the spatial arrangement
of capture events which are in fact highly informative about
individuality. Previous other treatments of uncertain identity in
capture-recapture models (Wright et al. 2009, Link et al. 2010, etc..)
ignored spatial information that is inherent in essentially all
capture-recapture studies, especially camera trapping studies.
Intuitively, and as I showed in the  example, the
configuration of encounters strongly affects the posterior
probabilities of individual association. Chandler and Royle (2013)
also deal with a type of SCR model that accommodates uncertain ID. In
their model they assume that individual detections are not
identifiable to individual (or
a subset is perfectly identifiable)
and they formulate the model in terms of a completely latent set of
$N$ encounter histories. In the present problem, we can 
construct either left-side or right-side encounter histories but are unable to
link the two data sets. The sample-based
uncertainty as in Chandler and Royle (2013), instead of
individual-based uncertainty considered here, may  be accommodated
in the present model which will be elaborated on in the sequel. 

Simulation results show that the model which accommodates unknown ID
typically performs well in terms of posterior mode (being centered
near the data generating value), and has small frequentist bias and
near the nominal 95\% frequentist interval coverage.  
In all cases of the proposed
estimator considered, the overall posterior mode was within 1
individual of the true population size, indicating that the posterior
is centered over the data-generating value. 
Results showed that there are increases in frequentist and posterior
precision as the number of known ID individuals increases (from 0 to
10 to 'all'). 
 Compared to the
heuristic estimator, which falsely assumes left and right side data
sets are independent samples of the population, the model which
accommodates unknown ID typically out-performs slightly in terms of
frequentist standard
deviation of the posterior mean and mode.
 The heuristic estimator is too liberal, under-states
posterior standard deviation, and produces confidence
intervals with less than nominal 95\% coverage. 
 This is to be expected since the heuristic estimator
over-states the sample size, counting left-side and right-side
encounter individuals as independent data sets.

The model proposed here accommodates the spatial arrangement of samplers but does
not require that the two sampling methods be co-located (for an
example see Gopalaswamy et al. 2012a).  Whether
samplers are co-located or not changes the possible observation states
only. For example, if single camera traps are used and DNA sampling
based on quadrat searches for scat are the 2nd method, and the
coordinates are not coincident, then $p^{camera}=0$ for camera trapping at
the DNA sites and $p^{DNA} = 0$ for DNA at the camera trapping sites.


The model proposed here seems to have direct relevance to the case of DNA sampling
when there are partial genotypes. For example, if a sample has 8 out
of 10 loci determined, with 2 missing, then the true genotype should be
regarded as latent and the model proposed here, with some
modifications, will integrate the spatial encounter information with
data on the observed loci, to give a probabilist match with
available genotypes in the vicinity (including possibly new genotypes
in the augmented population). A related problem is that of genotyping
error where, for any particular locus, there is some probability that
the observed allele is in error.  This problem is particularly
interesting because genotyping error has received considerable
attention recently (Lukacs and Burnham 2005; Wright et al. 2009) but,
as in our motivating camera trapping problem here, the spatial
information about individuality has not previously been incorporated
into existing methods that deal with genotyping error.  As in camera
trapping, when individuality is obtained from DNA sampling such as
from hair snares or scat obtained by dog searches, the spatial
location of captures should also be informative about individuality.
Thus, the basic idea developed here should be relevant to genotyping
uncertainty and also dealing with partial genotypes.

\section*{Acknowledgments}

Part of this research was performed using the ATLAS HPC Cluster, a
compute cluster with 672 cores, 4 Tesla M2090 GPU accelerators,
supported by  NSF grants (Award \# 1059284 and 0832782).

\section*{Literature Cited}

\newcommand{\rf}{\vskip .1in\par\sloppy\hangindent=1pc\hangafter=1                  \noindent}

\rf Bonner, S.J. and J. Holmberg. 2013. Mark-Recapture with Multiple,
Non-Invasive Marks. {\it Biometrics} 69, 766-775.

\rf Borchers, D.L. and M.G. Efford. 2008.
Spatially explicit maximum likelihood methods for capture-recapture studies.
{\it Biometrics} 64, 377-385.

\rf Chandler, R. B., and J. A.  Royle. 2013. Spatially explicit models for
inference about density in unmarked or partially marked
populations. {\it The Annals of Applied Statistics} 7(2), 936-954.

\rf Chandler, R. B. and J. D. Clark. 2014. Spatially explicit integrated
population models. {\it Methods in Ecology and Evolution} 5(12), 1351-1360.

\rf Efford, M. 2004. Density estimation in live-trapping studies. 
{\it Oikos}  106, 598-610.

\rf Gopalaswamy, A. M., Royle, J. A., Delampady, M., Nichols, J. D.,
Karanth, K. U., and  Macdonald, D. W. 2012a. Density estimation in tiger
populations: combining information for strong inference. {\it Ecology}
93(7), 1741-1751.

\rf Gopalaswamy, A. M., Royle, J. A., Hines, J. E., Singh, P., Jathanna,
D., Kumar, N., and K. U. Karanth. 2012b. Program SPACECAP: software for
estimating animal density using spatially explicit capture-recapture
models. {\it Methods in Ecology and Evolution}  3(6), 1067-1072.

\rf Karanth, K. U. 1995. Estimating tiger Panthera tigris populations
from camera-trap data using capture-recapture models. {\it Biological
Conservation}  71(3), 333-338.

\rf Link, W. A., Yoshizaki, J., Bailey, L. L., and K. H.
Pollock. 2010. 
Uncovering a latent multinomial: analysis of
mark-recapture data with misidentification. {\it Biometrics} 
  66(1), 178-185.

\rf Lukacs, P. M. and K. P. Burnham. 2005.  Review of capture-recapture
methods applicable to noninvasive genetic sampling. {\it Molecular Ecology}
14(13), 3909-3919.

\rf Magoun, A. J.,  D. Pedersen, C. Long and P. Valkenburg. 2011a.
Wolverine Images. Self-published at Blurb.com. 

\rf Magoun, A. J., C. D. Long, M. K. Schwartz, K. L. Pilgrim, R. E. 
Lowell and P. Valkenburg. 2011b. 
Integrating motion-detection cameras and hair snags for wolverine
identification. {\it The Journal of wildlife management} 75(3), 731-739.

\rf McClintock, B. T., P. B. Conn, R. S. Alonso, and K. R. Crooks.
2013. Integrated modeling of bilateral photo-identification
data in mark-recapture analyses. {\it Ecology} 94(7), 1464-1471.

\rf McClintock, B. T., L. L. Bailey, B. P. Dreher and
W.A. Link. 2014. 
Probit models for capture-recapture data subject to imperfect
detection, individual heterogeneity and misidentification. {\it The Annals of Applied Statistics} 8(4), 2461-2484.

\rf O'Connell, A. F., J. D. Nichols and  K. U. Karanth
(Eds.). 2010. 
 Camera traps in animal ecology: methods and
analyses. Springer Science \& Business Media.

\rf R Development Core Team. 2004.
R: A language and environment for statistical computing.
R Foundation for Statistical Computing Vienna, Austria.

\rf Royle, J. A., K.U. Karanth, A. M. Gopalaswamy and N.  S.
Kumar. 2009. 
Bayesian inference in camera trapping studies for a class of spatial
capture-recapture models. {\it Ecology} 90(11), 3233-3244.

\rf Royle, J. A., R. B. Chandler, C. C. Sun, and
A. K. Fuller. 2013. Integrating resource selection information with
spatial capture-recapture. {\it Methods in Ecology and Evolution} 4(6), 520-530. 

\rf Royle, J. A., R. B. Chandler, R. Sollmann and B. 
Gardner. 2014. 
 Spatial Capture-Recapture. Academic Press.

\rf Royle, J. A. 2009. Analysis of capture-recapture models with
individual covariates using data augmentation. {\it Biometrics} 
  65(1), 267-274.

\rf Royle, J.A., R.M. Dorazio and W.A. Link. 2007. Analysis of multinomial 
models with unknown index using data augmentation. 
{\it Journal of Computational and Graphical Statistics}  16, 67-85.


\rf Royle, J. A. and R. M. Dorazio. 2012. Parameter-expanded data
augmentation for Bayesian analysis of capture-recapture
models. {\it Journal of Ornithology} 152(2), 521-537.

\rf  Royle, J.A. and K.V. Young. 2008. 
A hierarchical model for spatial capture-recapture data.
{\it Ecology}  89, 2281-2289.

\rf Sollmann, R., Gardner, B., Parsons, A. W., Stocking, J. J.,
McClintock, B. T., Simons, T. R., Pollock, K.H and
A. F. O'Connell. 2013a. 
A spatial mark-resight model augmented with telemetry data. {\it Ecology}
94(3), 553-559.

\rf Sollmann, R., Gardner, B., Chandler, R. B., Shindle, D. B., Onorato,
D. P., Royle, J. A., and A. F. O'Connell. 2013b. Using multiple data
sources provides density estimates for endangered Florida
panther. {\it Journal of Applied Ecology} 50(4), 961-968.

\rf Sutherland, C., A. K. Fuller and J. A. Royle. 2015. Modelling
non-Euclidean movement and landscape connectivity in highly structured
ecological networks. {\it Methods in Ecology and Evolution}  6, 169-177.

\rf Wright, J. A., R. J. Barker, M. R. Schofield, A. C. Frantz, A. E. 
Byrom and D. M. Gleeson. 2009.  Incorporating Genotype
Uncertainty into Mark-Recapture-Type Models For Estimating Abundance
Using DNA Samples. {\it Biometrics} 65(3), 833-840.

\end{document}